\documentclass[conference, 10pt]{IEEEtran}
\IEEEoverridecommandlockouts
\usepackage[none]{hyphenat}
\usepackage{cite}
\usepackage{amsmath,amssymb,amsfonts}

\usepackage{graphicx}
\usepackage{textcomp}
\usepackage{xcolor}
\usepackage{comment}
\usepackage{url}
\usepackage{algorithmicx}  
\usepackage{algpseudocode} 
\usepackage{algorithm}
\usepackage{proof}
\usepackage{caption}
\usepackage{subcaption}
\usepackage{xcolor}
\usepackage[colorlinks,linkcolor=red,anchorcolor=blue,citecolor=green]{hyperref} 
\newcommand{\reffig}[1]{Fig \ref{#1}}

\newcommand{\refeq}[1]{Eq \ref{#1}}

\def\BibTeX{{\rm B\kern-.05em{\sc i\kern-.025em b}\kern-.08em
    T\kern-.1667em\lower.7ex\hbox{E}\kern-.125emX}}

\begin{document}
\author{\IEEEauthorblockN{Xiaoke Zhu\IEEEauthorrefmark{4}\IEEEauthorrefmark{1},
Qi Zhang\IEEEauthorrefmark{2}\IEEEauthorrefmark{1}\thanks{\IEEEauthorrefmark{1} made equal contribution to this work.}, Taining Cheng\IEEEauthorrefmark{4}, Ling Liu\IEEEauthorrefmark{3}, WeiZhou\IEEEauthorrefmark{4}\IEEEauthorrefmark{7}\thanks{\IEEEauthorrefmark{7} to whom correspondence should  address \{zwei,hejing\}@ynu.edu.cn } and
Jing He\IEEEauthorrefmark{4}\IEEEauthorrefmark{7}}
\IEEEauthorblockA{
Yunnan University\IEEEauthorrefmark{4},
%Beihang University\IEEEauthorrefmark{5},
IBM Thomas J. Watson Research\IEEEauthorrefmark{2},
Georgia Institute of Technology\IEEEauthorrefmark{3}\\
}}

\title{DLB: Deep Learning Based Load Balancing
	\thanks{This work was supported in part by the National Natural Science Foundation of China under Grant 61762089, Grant 61663047, Grant 61863036, Grant 61762092 and Open Foundation  of Key Laboratory in Software Engineering of Yunnan Province under Grand 2020SE310}
}
\maketitle
%\thispagestyle{fancy} % IEEE模板在\maketitle后会自动声明\thispagestyle{plain}，
% 导致第一页什么都没有。所以得把plain更改为fancy

%\pagestyle{fancy}
%\cfoot{\thepage}

%\rfoot{\thepage}

\begin{abstract}
	%Load balancing has been widely adopted by distributed platforms and its effectiveness is of great importance to the quality of services provided by such platforms. Hash functions are the core building block of the existing load balancing mechanisms. For example, Consistent Hashing is designed with the assumption that a hash function can uniformly map both client workloads and the servers to a circle. Thus with each workload being assigned to its clockwise closest server, the load distribution among different servers can be balanced. However, hash functions do not perform well on skewed data sets, which commonly exist in the real-world. Therefore, using hash functions in load balancing can lead to unbalanced load distribution and eventually harm the performance of applications and services running on the distributed platforms.
	
	In this paper, we introduce DLB, a \textbf{D}eep \textbf{L}earning based load \textbf{B}alancing mechanism, to effectively address the data skew problem. The key idea of DLB is to replace hash functions in the load balancing mechanisms with deep learning models, which are trained to be able to map different distributions of workloads and data to the servers in a uniformed manner. We implemented DLB and deployed it on a practical Cloud environment using CloudSim. Experimental results using both synthetic and real-world data sets show that compared with traditional hash function based load balancing methods, DLB is able to achieve more balanced mappings, especially when the workload is highly skewed.
\end{abstract}

\section{Introduction}	
With the development of Cloud computing, companies are becoming increasingly interested in migrating their services and data to Cloud platforms, such as AWS\cite{amazon2015amazon}, IBM Cloud \cite{ibmcloud}, Google Cloud \cite{googlecloud}, and Microsoft Azure \cite{msazure}, on which the effectiveness of load balancing is of great importance. Maintaining a balanced workloads benefits the cloud service provider by not only increasing the utilization of their resources, but also improving the quality of services. Currently, hash function based load balancing mechanisms are the dominant design, in which hash function based approaches are used to determine which server a workload needs to be assigned to. %\footnote{To simplify the illustration, we may also use a ball to represent a client workload, and a bin to refer to a server.}.%in which the input data points are usually represented as balls, and the target servers are represented as bins. \footnote{Throughout this paper we use a ball to represent a client workload, and a bin to refer to a server.}. Given the ID of a ball, a hash function based load balancing mechanism is designed to find a bin that this ball should be assigned to. Meanwhile, both balls and bins can be added or removed dynamically.

A hash function is able to generate balanced results when the input data is uniformly distributed. However, the real-world data sets often exhibit remarkable skew. For instance, analysis of air traffics and online human behavior data sets\cite{jiang2011exploring,radicchi2009human} revealed that such data usually follows different power law distributions. When the input data is skewed, the output of the hash function will also be skewed. Therefore, using a hash function in a load balancing mechanism can result in unbalanced workloads assignment when data skew exists in the input. Even worse, such unbalanced workloads could seriously harm the performance of applications and services running on distributed platforms. Elaheh Gavagsaz and et al. \cite{DBLP:journals/tjs/GavagsazRJ19} demonstrated that traditional join algorithms based on MapReduce are not efficient when working with skew data, Joanna Berlinska and et al. \cite{DBLP:journals/pc/BerlinskaD18} also revealed that the uneven distribution of the keys might cause imbalance computation completion time among different MapReduce tasks, which eventually prolonged execution of the whole MapReduce job.

\begin{figure}[t]
	\begin{minipage}[t]{1.0\linewidth}
		\centering
		\includegraphics[width=0.8\linewidth]{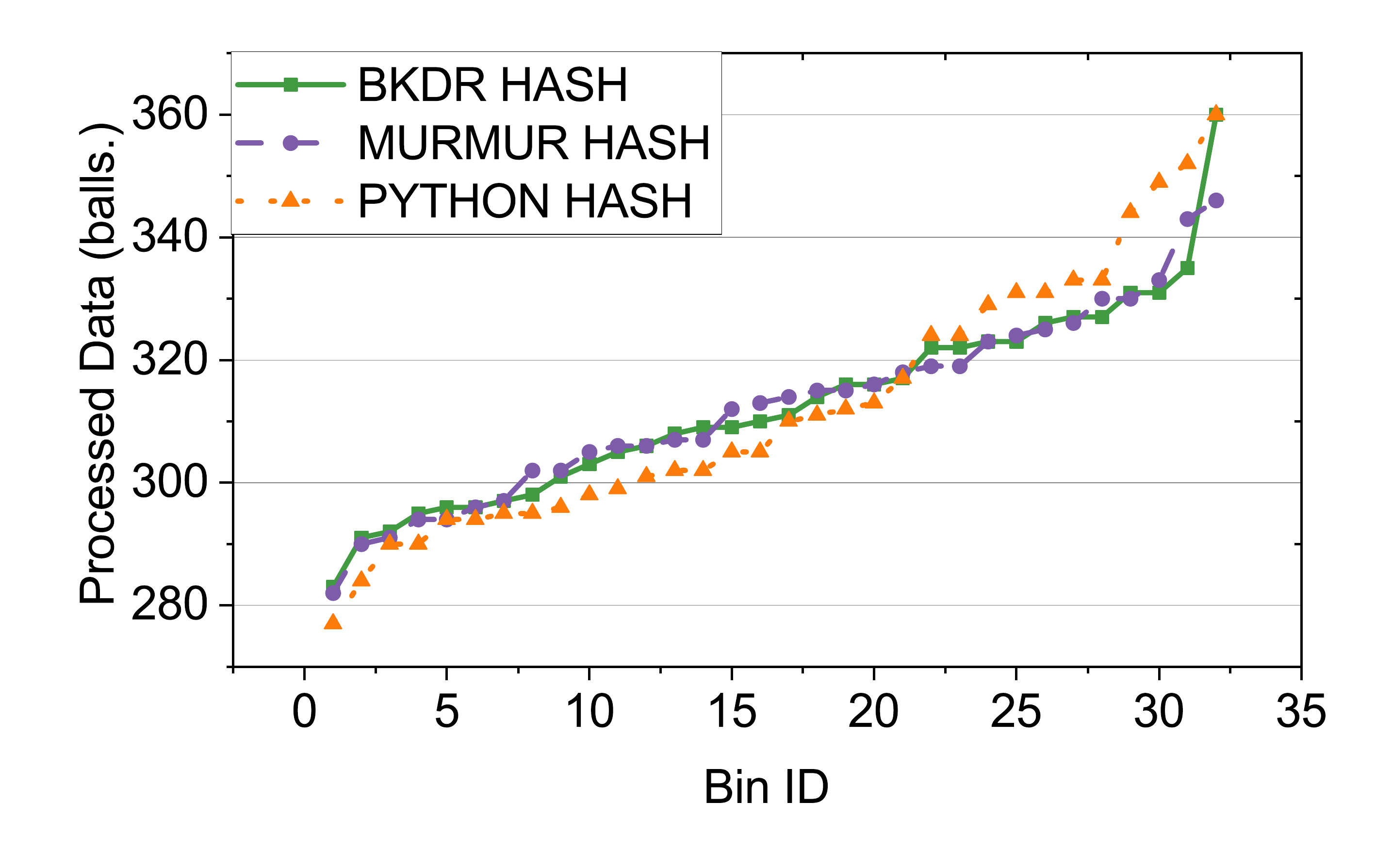}
		\caption{Data skew for different hash functions}
		\label{fig:data-skew}
	\end{minipage}
\end{figure}

There are several reasons why hash functions do not perform well on skewed data sets. First, the hash function was originally designed to perform fast index \cite{ThomasH2002Introduction} (i.e., indexing with $O(1)$ time complexity), compression \cite{DBLP:conf/icisc/HiroseKY11} (i.e., compressing a large input in a deterministic way), cryptography \cite{Coles2009Asymmetric} (i.e., irreversible mapping from inputs to outputs) and etc., thus dealing with skewed data was not considered as one of its primary design goals. Second, although there have been efforts, such as BKDR hash \cite{BKDR-hash}, MURMUR hash \cite{MURMUR-hash}, and Python hash \cite{Python-hash}, to enhance the hash functions to better handle the skewed data, their effectiveness are not satisfying. 
%For example, BKDR HASH \cite{BKDR-hash} allows each element of the input data to participate in the calculation of hash to cause an avalanche effect. In this way, even with small changes to the input data, the output of BKDR can change drastically. In addition to letting every element of the input data participate in the calculation, MURMUR HASH \cite{MURMUR-hash} uses shift and subtraction instead of multiplication. Thus the output of MURMUR HASH has a high balance and low collision rate. However, the results are still not satisfying. 
%\begin{comment}
As shown in Figure \ref{fig:data-skew}, we manually generate a skew data set under normal distribution as the inputs and use the above mentioned three hash functions to map these inputs to 32 bins. After that, the bins are sorted by the number of inputs assigned to it. Ideally, the lines in this figure should be flat, which means different bins are taking a similar amount of inputs if the hash functions are able to map the input to a uniformed distribution. However, the lines in the figure are all in an increasing trend. The numbers of inputs being assigned to different bins vary from 280 to 360, which are significantly unbalanced.

%\end{comment}
The availability of big data and the rapid advance of AI techniques provide unique opportunities to rethink the design of load balancing mechanisms by making them perform better on skew data. The key idea is as follows: instead of using a hash function, learned models can be applied to determine where the inputs should be mapped to the hash circle. %Such learned models are trained on historical data sets, which are appropriately labeled to avoid the data skew problem. Then, these models can be used to effectively map the newly coming data with similar distributions into a uniformly distributed space. 
Although model training is required beforehand, this approach is practical and has several advantages compared to traditional hash based methods. First, the capability and affordability of collecting large amount of data nowadays provides the potential to train data distribution aware models for load balancing mechanisms.
%collecting larger amount of data for analytic purpose is becoming the trend for many business to better serve their customers, thus such data can also be used to train data distribution aware models for load balancing mechanisms. 
Second, the distribution of data collected by a specific company or organization for a given task is usually consistent, which is demonstrated by analysis results from \cite{jiang2011exploring,brockmann2006scaling,radicchi2009human}. This shows the feasibility of using historical information to deal with future workloads. Third, when appropriately trained, the output of a learned model can be uniformly distributed even when the input data set is highly skewed.

In this paper, we propose DLB, which uses deep learning models to effectively address the data skew problem in existing load balancing mechanisms. Researchers \cite{DBLP:conf/sigmod/KraskaBCDP18,DBLP:journals/access/XiangZCCLZ19,DBLP:conf/cidr/KraskaABCKLMMN19,DBLP:conf/sigmod/GalakatosMBFK19} have explored the possibility of partially replacing existing data structures and algorithms with deep learning models. For example, Tim Kraska and et al. \cite{DBLP:conf/sigmod/KraskaBCDP18} introduced the hash model index, which reduces the total number of hash conflicts over map data set by learning a CDF(Cumulative distribution function) at a reasonable cost.
%Moreover, an LSTM(Long Short Term Memory) based inverted index was developed by Wenkun Xiang and et al. \cite{DBLP:journals/access/XiangZCCLZ19}. It is able to achieve lower average search time compared with traditional inverted index structure by taking the advantages of the learned model. 
However, there are still remaining challenges to leverage deep learning models to improve the effectiveness of load balancing mechanisms. On the one hand, how to design a neural network that can converge quickly during the training while also being able to effectively mapping large volumes of inputs to a uniformly distributed space. On the other hand, how to balance between the complexity and the expressiveness of the model. Concretely speaking, a simple neural network can be easily trained, but it will not be able to map large amount of inputs into a uniformly distributed space without incurring significant conflicts. While a complex model can reduce the mapping conflicts, but it cannot be trained easily due to gradient dissipation and explosion problems. %Another challenge is how to adjust the other components {\color{orange} (need to be more specific on what components)} in the existing load balancing mechanisms so that they can collaborate well with the deep learning model to provide balanced mapping results, especially in a dynamic environment where servers can dynamically join and leave.

In order to solve these challenges, DLB is designed in a way that, instead of using a single end-to-end model, it organizes a set of models into a hierarchical architecture. In such an architecture, the models are organized in different connected layers. For a specific input, it will go through one model in each layer, while the model in the previous layer specifies which model in the next layer needs to be invoked. The final output will be the position on the hash circle for this input. Since the distribution of input data is not guaranteed to stay the same, DLB also continuously monitors the actual load distribution of all the servers to make sure no server becomes a hotspot. %, otherwise, the workload will be assigned to a different server with less load.
Compared with traditional hash function based load balancing mechanisms, such as Consistent Hashing\cite{DBLP:conf/stoc/KargerLLPLL97} and Consistent Hash with Bound Load \cite{DBLP:conf/soda/MirrokniTZ18}, DLB is able to map the input data sets to a uniformly distributed space even when they are highly skewed. In addition, compared with a single but complex end-to-end model, a hierarchical design makes each model converge more quickly during the training. %Experiments over both synthetic and real-world data sets with different distributions demonstrated that, compared with traditional hash function based load balancing methods, DLB is able to generate more balanced results when the input data is skewed.

The main contributions of this paper are as follows:
\begin{itemize}
%	\item We investigated the data skew problem in traditional load balancing mechanisms and explored the opportunities to improve their effectiveness by using deep learning models.
	\item We designed DLB, a deep learning based load balancing mechanism which solves the data skew problem by replacing the hash function with deep learning models. %(\refsec{sec:framework})
	\item We implemented DLB and deployed it in a practical environment using CloudSim \cite{calheiros2011cloudsim}, which enables modeling and simulation of real Cloud computing systems and application provisioning environments. 
	\item The effectiveness of DLB is measured by using both synthetic and real-world data sets under different distributions.
\end{itemize}

%The rest of the paper is organized as follows: the related work is introduced in \refsec{sec:related-work}. \refsec{sec:framework} discusses the design of DLB. The experiment setup is described and the results are analyzed in \refsec{sec:experiments-design}. \refsec{sec:conclusion} gives the conclusions and future directions.

\section{Related Work}\label{sec:related-work}
Load balancing mechanisms are widely used in distributed computing environment to balance the workloads among different servers, and the effectiveness of such mechanisms is critical to the overall performance and service quality of the distributed platforms. Therefore, how to design an effective load balancing mechanism has attracted the interest of many researchers. In this section, we introduce related researches in this area, while at the same time, we also discuss the existing efforts on trying to use neural network based learned data structures to improve the performance of traditional systems.

\textbf{Hashing based load balancing.} As one of the mainstream load balancing mechanisms, Consistent Hashing(CH) \cite{DBLP:journals/mst/KargerR06} proposed by Karger and et al. has been widely adopted. Ideally, by using a randomized hash function, both balls and bins can be assigned to the hash circle in a uniformed way, so that different bins will be able to hold the similar number of balls. However, it is usually not the case in the real-world due to the existence of data skew in the inputs. There have been many efforts to address this issue \cite{DBLP:journals/ton/ThalerR98,DBLP:journals/corr/LampingV14,DBLP:conf/esa/GrossiV18,DBLP:journals/monet/WangR09}. For example,  David R. Karger and el al. \cite{DBLP:conf/stoc/KargerLLPLL97} tried to enable CH to generate more balanced results by using virtual bins which are replicas of real bins in hash space, and one real host can be correspondent to several virtual bins. The authors showed that the overall load balancing performance could be improved accordingly. To further address the data skew problem in load balancing, Johan Lamping and Eric Veach proposed jump Consistent Hashing\cite{DBLP:journals/corr/LampingV14}, which works by computing when its output changes as the number of bins varies. In this approach, the hash value of a ball is not randomly generated, but acquired according to the probability determined by the number of existing bins. Also, whenever a new ball is added, the hash value of the existing balls needs to be recomputed according to the a predefined probability. Roberto Grossi and et al. designed Round-Hashing \cite{DBLP:conf/esa/GrossiV18}. Thaler and Ravishankar proposed \cite{DBLP:journals/ton/ThalerR98} Rendezvous hashing algorithm, for a given ball $q$ and $n$ bins, it applies a hash function to the all the pairs $\{q,p_i\}$, in which $i \in \{1...n\}$, and assigns the ball to the bin that can lead to largest hashing result. %However, this process is time consuming since it needs to rehash the pair of $q$ and all the balls when adding a ball $q$.

\textbf{Neural network based learned data structures.} This thread of research explores the potential of utilizing the neural network based learned data structures to improve the performance of traditional systems \cite{nn-sort,DBLP:conf/sigmod/GalakatosMBFK19,DBLP:conf/sigmod/KraskaBCDP18,DBLP:journals/access/XiangZCCLZ19}. Among them, Tim Kraska and et al. proposed learned B-tree, learned hash, and learned bloom filter structure \cite {DBLP:conf/sigmod/KraskaBCDP18} to improve the indexing performance of traditional structure by learning the distribution from the historical data. Xiang and et al. \cite{DBLP:journals/access/XiangZCCLZ19} proposed a LSTM-based inverted index structure.%, in which a learned inverted index structure led to fewer average look-ups compare to tradition inverted index structures. %Alex Galakatos et al. \cite{DBLP:conf/sigmod/GalakatosMBFK19} presented a data-aware index structure named FITing-Tree, which approximates an index using piece-wise linear functions with a bounded error specified at construction time. \cite{nn-sort}

Different from the above-mentioned researches, instead of improving the effectiveness of traditional hash functions, our DLB approach takes the advantages of both Consistent Hashing and deep neural network. In DLB, a deep learning model based on the historical data is trained, and then used to map the newly coming data to a uniformly distributed space even when such data is skewed.
\begin{figure*}[t]
	\begin{minipage}[t]{1.0\linewidth}
		\centering
		\includegraphics[width=0.7\linewidth]{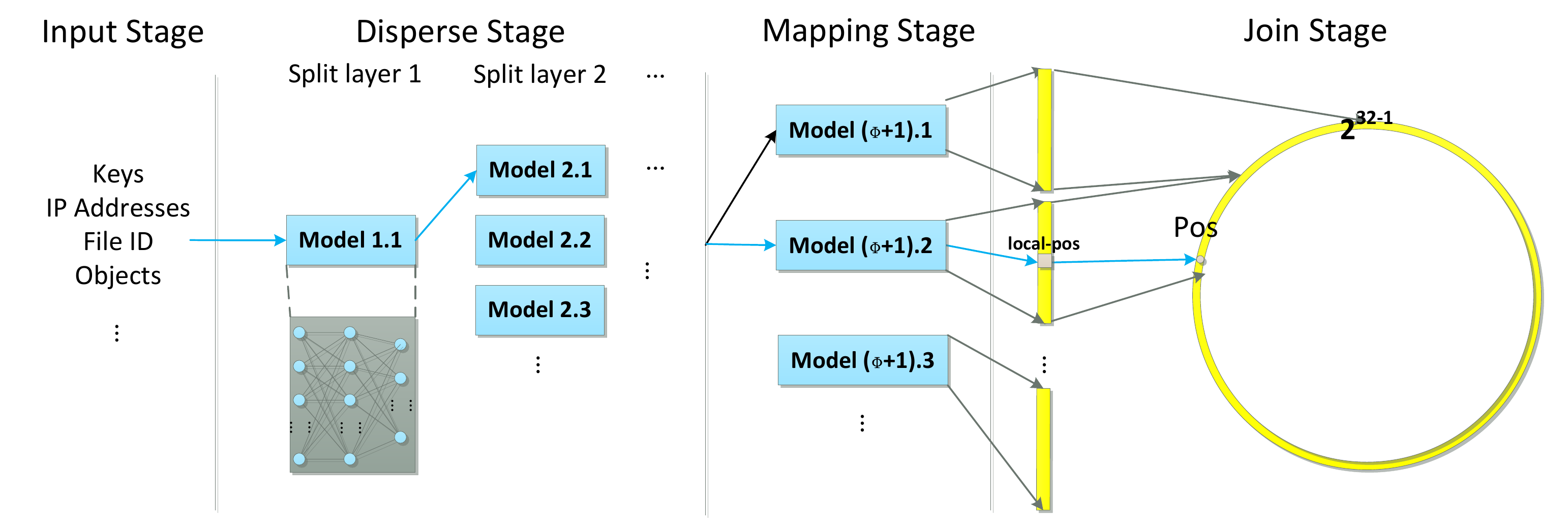}
		\caption{The hierarchical model architecture in DLB}
		\label{fig:framework}
	\end{minipage}
\end{figure*}
\section{DLB: Deep learning based load balancing}\label{sec:framework}

In this section, we discuss the details of DLB design. The goal of load balancing is to uniformly distribute different workloads on multiple servers so that no server will become the hotspot. A hash function, being the core building block in most of the existing load balancing mechanisms, can be considered as a black box that takes an input and maps it to a position on the hash circle. %Considering the fact that the most critical task is to find a mapping between the balls and their positions on the hash circle, whether it is accomplished by a hash function or not is not important. 
Therefore, we propose a \underline{D}eep \underline{L}earning based load \underline{B}alancing mechanism named DLB, which replaces the hash functions with deep learning models to fulfill the same mapping task. We observe that this approach is able to work well on skew data and provide more balanced workload distribution compared with traditional hash function based load balancing mechanisms.

\subsection{Design}
\subsubsection{Hierarchical models}\label{sec:kernel}
As discussed earlier, a hash function in a load balancing mechanism can be replaced by a deep learning model, and a well trained model can generate uniformly distributed outputs even when the inputs are highly skewed. A natural question to ask is which model should be used. In load balancing mechanisms such as CH, the inputs are usually mapped into a large space (e.g., $2^{32}$) to avoid conflicts. If we consider each slot in this output space as a class, what the model needs to achieve is actually classifying each input into one of these different classes. This actually turns the mapping task into a classification task. Since the space of this classification so large, training a single model for such a task will be really difficult.

Therefore, we propose an architecture of hierarchical models in DLB to address this problem. As shown in Figure \ref{fig:framework}, instead of using one single model, multiple models are involved to solve this classification problem. The models are organized into a hierarchical structure with different connected layers. To find out the position that an input should be mapped to on the hash circle, each input will need to go through a model in each layer. The whole mapping procedure can be divided into 4 stages, and details of each stage are described as follows:

\textbf{Input stage.} %Various tasks makes difference input to load balancing approaches. For instance, distributed database system may take key value as the input, some cloud file systems take MD5 code as input. Then the question is naturally arisen, how to unify the input format and make learned model can dear with it.  
Various formatted features can be observed as the inputs of load balancing mechanisms when a hash function is used, being it an ID string of a user or a MD5 value of a file. However, these features need to be converted so that they can be consumed by a neural network model. Therefore, the goal of this stage is to pre-process the input data, such as converting strings or numerical data into vectors, so that they can be directly used as inputs of a neural network model. 

\textbf{Disperse stage.} The main strategy used in this stage is \textit{divide and conquer}. Concretely speaking, the disperse stage consists of multiple models which are organized in a hierarchical architecture(i.e., a tree structure). All the models in this architecture work collaboratively to figure out which position on the hash circle a given input should be placed. Since the space of the final hash circle is usually very large, the motivation of this design is to \textit{divide} a complex classification task, which is supposed to deal with a large output space, into multiple smaller tasks. In this way, the original complex classification problem can be \textit{conquered} more effectively by solving these smaller problems. In other words, with such a hierarchical architecture, each model only needs to tackle a much simpler classification problem with only a subset of the whole output space. As shown in Figure \ref{fig:framework}, the models in this stage are split into multiple layers. The root model(i.e. the one on the left most in Figure \ref{fig:framework}) takes the input data set and determines which model in the next layer needs to be invoked, while the models in the other layers of the disperse stage go through the same procedure using their corresponding assigned input.

\textbf{Mapping stage.}
\begin{comment}
\begin{figure}[t]
	\begin{minipage}[t]{1.0\linewidth}
		\centering
		\includegraphics[width=3.in]{tensorflow-output.pdf}
		\caption{The output structure of mapping stag model}
		\label{fig:output}
	\end{minipage}
\end{figure}
\end{comment}
Models involved in this stage are located in the last layer of the hierarchical architecture. Different from the models in the disperse stage that select which model in the next layer needs to be used, each model in the mapping stage is responsible for generating the position of a given input in its sub-circle. Different models in the mapping stage are correspondent to different sub-circles, which are actually areas on the hash circle, while they collectively cover the whole hash circle. %As shown in  \reffig{fig:output}, the models in this stage generates a $logits$ vector of size $t$ for ball, in which $t$ equals to the size of sub-circle this model is responsible for. For a ball $q_i$, its corresponding output vector $v_i$ produced by the model in the mapping stage needs to go through the softmax and argmax functions to generate the local position on the sub-circle. 

\textbf{Join stage.} Since the output of each model in the mapping stage is a position on each model's own sub-circle, another layer is needed to translate such a local position on a sub-circle into a global position on the hash circle. In order to create the continue hash circle, sub-circle of different mapping stage models are connected sequentially, thus the final global position $Pos_{qi}$ of input $q_i$ on the hash circle is established by \refeq{eq:pos}.
\begin{equation}\label{eq:pos}
Pos_{q_i} = \mu_i+ (model\_id - 1)t
\end{equation}

 $model\_id$ is the ID of the model in the mapping stage starting from 1.

\subsubsection{Server Management}\label{sec:dynamic-adjust-rules}
%Simply replacing the hash function with hierarchical model is not enough for existing loading balancing mechanisms such as CH. The deep learning load balancing mechanism behavior {\color{orange} what does this mean?} regarding to the dynamic addition and removal of the servers also need to be considered.

In traditional load balancing mechanisms such as CH, both workloads and servers are mapped to the hash circle by hash functions, and a workload is assigned to its clockwise closest server. In DLB, a deep learning model is used to map a workload to a position on the hash circle, while a deterministic approach is used to map the servers. Although server mapping can also be done by using learned models, it is not necessary since the number of servers is usually much smaller than that of the workloads and a deterministic approach is good enough to evenly map the servers to the hash circle.

Similar to traditional load balancing mechanisms, DLB assigns a workload to a server in a clockwise manner. Since DLB is able to uniformly map the workloads to the hash circle, using a deterministic server mapping approach can achieve well balanced workload distribution.
%The assumption is that the hash function can uniformly mapping both balls and bins to the hash circle, which is, however, not usually true in the real-world cases as we discussed earlier. 
%In DLB, the workloads can be uniformly mapped to the hash circle by well trained hierarchical models, while servers are even mapped to the hash circle in a deterministic way. Since the workloads are assigned to the servers in the same clockwise manner, this approach allows the load among different servers to be well balanced. 
Concretely speaking, when a new server is added, DLB will add the server to a place such that this server can evenly divided the largest sub-circle on the hash circle. %For example, the first server will be placed to a random position on the hash circle, while the second server should be in a position which can divide the circle into two halves with the first bin. When a third server comes, it will be placed to a position which equally divide either half of the circle into another half, so that with the fourth server, the hash circle can be divided equally into four partitions, so on so forth. 
When an existing server becomes unavailable, the workloads on this server will be reassigned to its clockwise next server. Similar to Consistent Hashing with bounded load(CHBL), each server in DLB has a load threshold $\epsilon$. A new workload can be assigned to a server only if the load of this server will not exceed $\epsilon$. Otherwise, other servers need to be considered. %Our experimental results show that this approach performs well to keep balanced workloads as well as to minimize the ball migration costs when the available servers dynamically changes. 

\subsection{Training}
In this sub-section, we discuss the considerations of how to train the hierarchical models mentioned above from two aspects. 

First, how to label the training data. Given historical cluster access data, to make sure that the models will not generate skew output %the labels of the training data need to be uniformly distributed 
when the training data is skewed. Meanwhile, %since DLB contains a hierarchical of models instead of one single model, the labels also need to be adjusted accordingly. 
Since the hierarchical architecture includes multiple models in different layers, for each input, it needs to be labeled for each model. We discuss the labeling process for DLB from two aspects: creating labels used in the mapping stage models as well as in the disperse stage models. The difference between these two types of labels is that the former one represents positions on a hash circle, while the latter one is correspondent to the ID of the model in the next layer.

\begin{algorithm}
	\centering
	\caption{ Labeling DLB training data}\label{algo}
	\begin{algorithmic}[1]
		\Require $K$ - key list of balls
		\Require $T$ - number of positions on the hash circle
		\Require $\Phi$ - number of layers in disperse stage
		\Require $labels^i$ - labels for the models in the $i$th layer
		\Require $indexof(k_i, K)$ - index of the element $k_i$ in list $K$
		\Require $tag^{\phi}(k_i)$ - label of $k_i$ for the model in $\phi$th layer.
		\Ensure $Labels^i$ $\gets$ $\{ \}(i\in (1, \Phi+1))$
		\State $K_s$ $\gets$ Sort($K$)
		
		\For{$k_i$ in  $K$} //Create labels for models in mapping stage
		\State $label$ $\gets$ $indexof(k_i, K)*(T/sizeof(K))$
		\State $labels^{\Phi+1}$ $\gets$ $Labels^{\Phi+1}$ $\cup$ $label$
		\EndFor
		
		\For{$\phi$ in $\Phi$} //Create labels for models in disperse stage
		\For{$k_i$ in  $K$}
		\State $label$ $\gets$ $tag^{\phi}(k_i)$
		\State $labels^\phi$ $\gets$ $Labels^\phi$ $\cup$ $label$
		\EndFor
		\EndFor\\
		\Return $\{labels^1, ...,labels^{\Phi+1}\}$
	\end{algorithmic}
	\label{alg:labeling}
\end{algorithm}

The method to create labels for DLB is depicted in Algorithm \ref{alg:labeling}. 
%The Algorithm first sorts the keys of all the training data(line 1). For each training data point, since it needs to go through one model in each layer during the inference, it has a label corresponding to each layer. How to create the labels for the models in the mapping stage(i.e., layer $\Phi+1$) is described by line2 - line5, in which the inputs are mapped to a hash circle in a uniformed manner. Line6 - line11 generate the labels for the disperse stage(i.e, layer1 - layer$\Phi$).
$tag^{\phi}(k_i)$ generates the label of $k_i$ in layer $\phi$. A formal description of $tag^{\phi}(k_i)$ is shown in Eq \ref{eq:label:disperse-stage}, in which $C_\phi$ represents the the number of models in the $\phi$th layer.

\begin{equation}\label{eq:label:disperse-stage}
tag^{\phi}(k_i)  = j \qquad j\in[1:C_{\phi+1}]
\end{equation}
\begin{displaymath}
\textrm{subject to:}\ \frac{T}{C_{\phi+1}}*(j) \le\ label^m_{i} \le\ \frac{T}{C_{\phi+1}}*(j+1)
\end{displaymath}

Second, we also describe what loss function is used in the training process. The loss value used in the training is defined as $Loss = \sum_{\phi \in (1, \Phi+1)}\sum_{n \in (1, m_\phi)}(O^n_\phi-label^n_\phi)^2$, which is the sum of the loss value of each model in the hierarchical architecture. We refer the output of $n$-th model in the $\phi$-layer  to as $O^n_\phi$, $label^n_\phi$ to represent the corresponding labels, while $m_\phi$ as the number of models in the $\phi$th layer.

%Considering the fact that a model can be trained once and used multiple times, and also each single model used by DLB is much simpler than those used in visual recognition and natural language processing, this paper does not focus on discussing the training overhead, and we put it as our future work.
 
%It should be pointed out that all of our models, shallow neural networks, train relatively fast and their interface time (predicting time) could be very tiny. Besides, the model can be used over and over again under tasks with similar data distribution.  Hence, The training  time is not the focus of this paper.
\begin{figure*}
\begin{subfigure}{1\textwidth}
\centering
\includegraphics[width=0.9\textwidth]{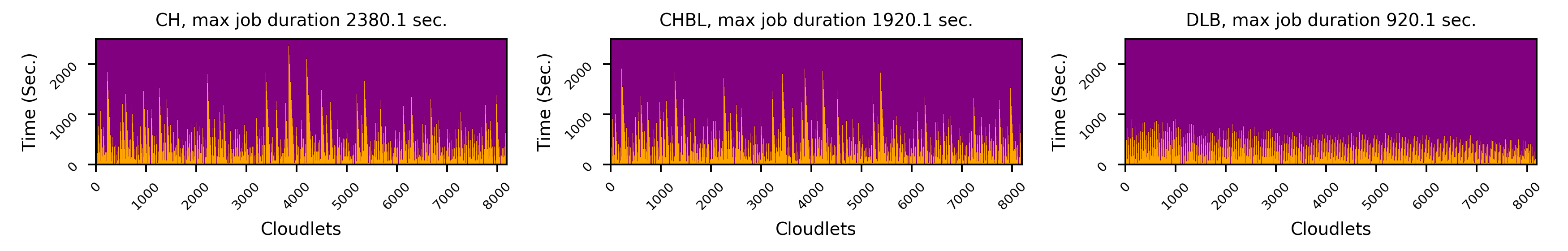}
\caption{Log-normal distribution}
\label{fig:analysis:cloudsim:log-normal}
\end{subfigure}
\begin{subfigure}{1\textwidth}
\centering
\includegraphics[width=0.9\textwidth]{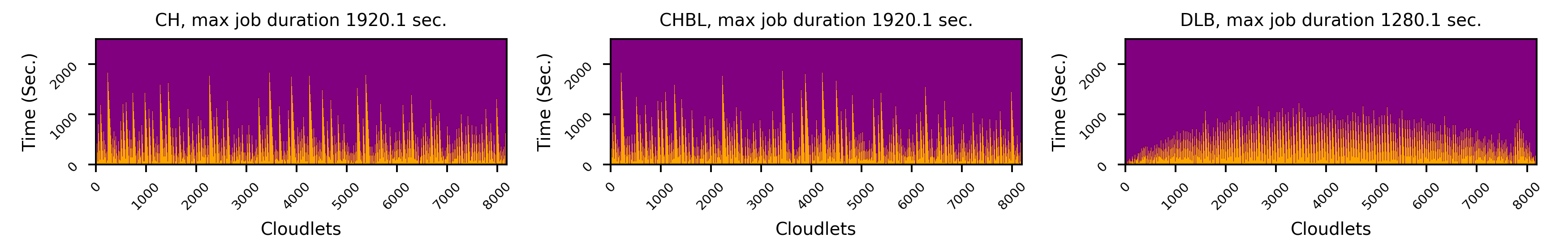}
\caption{Normal distribution}
\label{fig:analysis:cloudsim:normal}
\end{subfigure}
\begin{subfigure}{1\textwidth}
\centering
\includegraphics[width=0.9\textwidth]{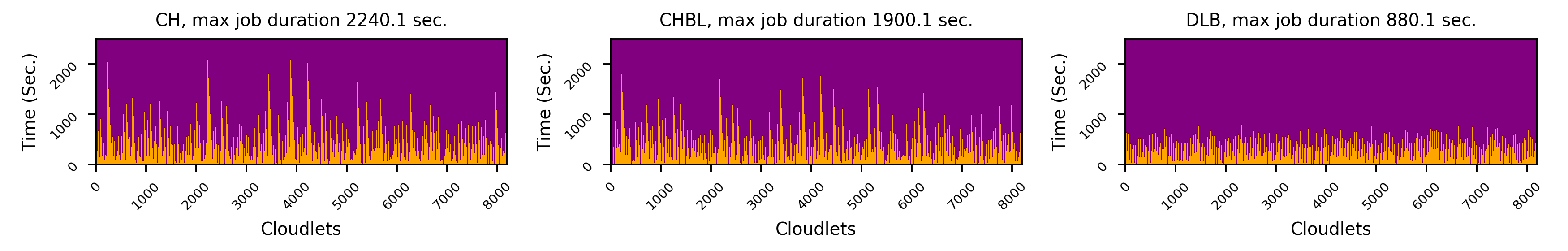}
\caption{Uniform distribution}
\label{fig:analysis:cloudsim:uniform}
\end{subfigure}
\caption{Compare the effectiveness of different load balancing mechanisms in a practical Cloud environment created by CloudSim.}
\label{fig:analysis:cloudsim}
\end{figure*}
\section{Evaluation}\label{sec:experiments-design}
In this section, we compare the effectiveness of load balancing between DLB and the following widely used and classical load balancing approaches:
\begin{itemize}
	\item Consistent Hashing(CH). CH hashes the balls and bins into a unit circle, and uses the hash values to create a circular order of balls and bins. The palcement decision are all based on the relatevie location among balls and bins.%Assuming no collisions, a ball is placed in its clockwise closest bin. If a bin is removed, CH moves the balls in this bin to its next clockwise closest bin. Similarly, when a new bin is added, only part the balls whose positions are between the new bin and its clockwise closest bin will need to be adjusted.
	\item Consistent Hashing with Bounded Load(CHBL). CHBL is similar to CH, except that it uses a parameter to try to keep the balls uniformly distributed among different bins.% Before assigning a ball to a bin, CHBL makes sure that the load of that bin will not exceeds some threshold after the ball is added. Otherwise, the ball will be assigned to another bin.
	%\item Rendezvous Hashing(HRWH). The basic idea of HRWH is to give each bin a weight for each ball via a hash function, and assign the ball to the highest weighted bin. Compared to CH, HRWH has a higher time complexity when adding bins. Specifically, the complexity of adding a bin to the hash circle in CH is $\hat{O}(\beta+logn)$, but it is $O(\beta n )$ in HRWH. $\beta$ is the load factor and $n$ is the number of existing bins \cite{DBLP:conf/esa/GrossiV18}.
\end{itemize}	
In our design each sub model of DLB has 3 fully connected layer, and each layer has 8, 32, 64 neuros respectively. We use Adam \cite{kingma2014adam} with learning rate of $0.01$ to train all the sub models. For CH and CHBL, we also combine them with different hash implementations, such as BKDR Hash \cite{BKDR-hash}, Python Hash \cite{Python-hash}, and Murmur Hash \cite{MURMUR-hash}. %Since all these hash functions are designed to improve the balance of data mapping, we compare them with the effectiveness of DLB.

\subsection{Setup}
In this subsection, we describe the experimental setup, including the hardware and software environment, as well as the data sets and metrics used throughout the measurements.

\subsubsection{Environment}
The experiments are carried on a machine with 64GB main memory and one 2.6GHZ Intel(R) i7 processors. Each test is run 10 times and the median of the results are shown in this section.

%Two GTX1080Ti GPU card are installed and each of them has 16G GPU memory. RedHat Enterprise Server 6.3 with Linux core 2.6.32 was installed, and we use Tensorflow  for experiments. Each test is run 10 times and the median of the results are shown in this section.

\subsubsection{Data sets}
In order to measure how different distributions of the input data set can affect the effectiveness of DLB, the synthetic data sets used in Section \ref{sec:analysis} are generated under three most commonly observed distributions: uniform distribution, normal distribution, and log-normal distribution. Each distribution has two data sets, one for testing and the other for training. Each data set consists of $20,000,000$ balls while each ball is a double-precision digit which represents a key of client workload in load balancing scenario.  A 4TB data set collected from a radio monitoring center is also used in our experiments. This data set consists of $100,000$ records, and each record has 13 features.

\subsubsection{Measurement metrics}
The effectiveness of a load balancing mechanism is measured by the standard deviation among the load of different bins on the hash circle(std). Therefore, the smaller the std value is, the more effective the load balancing mechanism is. Formally, the $std$ can be calculated as $std=\sqrt{\frac{\sum_{j=1}^{n}(load_{j}-\frac{m}{n})^2}{n}}$, in which, $load_{j}$ refers to the number of balls assigned to bin $j$, while $m$ and $n$ are the total number of balls and bins respectively. %Thus $\frac{m}{n}$ represents the load on each bin when the balls are uniformly distributed. Note that load balancing performance  and  accuracy of the hierarchy of the model reflect the same thing in this paper.

\subsection{Analysis}\label{sec:analysis}
\subsubsection{CloudSim based evaluation}\label{exp:cloudsim}

%The first tests that we present here are aimed at analyzing the max duration time of jobs on the cloud enviroment. This critical factor must be considered during the VM provisioning process by a Paas provider, to avoid unnecessary VM running overhead. While, it also impact the rental time and prices of PaaS users. That is, the more balancing allocating between VM and task requests from users, the less VM managing overhead for PaaS provider, and the cheaper rental price for users. The test simulation environment setup for measuring the max duration time by CloudSim\cite{calheiros2011cloudsim} which is  a toolkit for modeling and simulation of cloud computing environments. In our design, 64 machines were hosted within a single data center. Each machine hold 4 threads, 1 GB of RAM, 10 GB of storage, and its band width is set to 1GB/sec..  The VMs created in the public cloud were based on an Amazon’s small instance with a little adjusting (1 GB of memory, 1 virtual core, and 10 GB of instance storage). The workload,  subjected to real word common distributions, sent to the cloud was composed of 8192 tasks denoted as Cloudlets, in which each cloudlet should take about 20 sec to be finished (only affected by bandwidth, computing resources, and allocating policy).  We compare the duration time of DLB agatint two benchmark load balancing algorithm, CH and CHBL, as aforesaid. Noted that, Murmur Hash is set as kernel of the all benchmarks.

In this set of experiments, we deploy DLB, CH, and CHBL on a practical Cloud environment created by CloudSim \cite{calheiros2011cloudsim}, and compare their effectiveness to balance the workloads among a number of servers. In the simulated Cloud environment, 64 machines are hosted within a single data center. Each machine has 4 threads, 1 GB memory, 10 GB of storage, and 1Gbps network bandwidth. On the client side, 8192 workloads(cloudlets) with different distributions are created. The distribution is based on the ID of each workload, which is also the key used for mapping. Each server can run at most 4 workloads simultaneously. When wait list of a server is full and an additional workload is assigned, this workload needs to be loaded by the next spare server in the hash circle. All the workloads are submitted to the data center in one batch, and each workload takes 20 seconds CPU time to finish.

\reffig{fig:analysis:cloudsim} shows the actual finishing time of each workload when different load balancing mechanisms are used. The x-axis represents the ID of each workload while the y-axis shows the actual finishing time. It can be observed that, no matter what the workload distribution is, the heights of lines in figures corresponding to DLB is much lower and smoother. This means that, when DLB is used, workloads can finish in a shorter and more balanced time. While in CH and CHBL cases, some workloads takes much longer time to finish than the other ones. This is due to the imbalanced assignment of workloads, which causes some servers to become the hotspots. Therefore, it takes much longer for workloads on these servers to finish. Considering the exact running time of the longest job (i.e., max job duration) in each scenario, DLB is able to reduce such time by $61.3\%$ over CH and $52\%$ over CHBL under log-normal distribution, $33\%$ over both CH and CHBL under normal distribution,  $60.7\%$ and $53.68\%$ when compared with CH and CHBL respectively under uniformed distribution.

\begin{figure}%\label{fig:analysis:load-balance}
	\begin{subfigure}{.22\textwidth}
		\centering
		\includegraphics[width=1\linewidth]{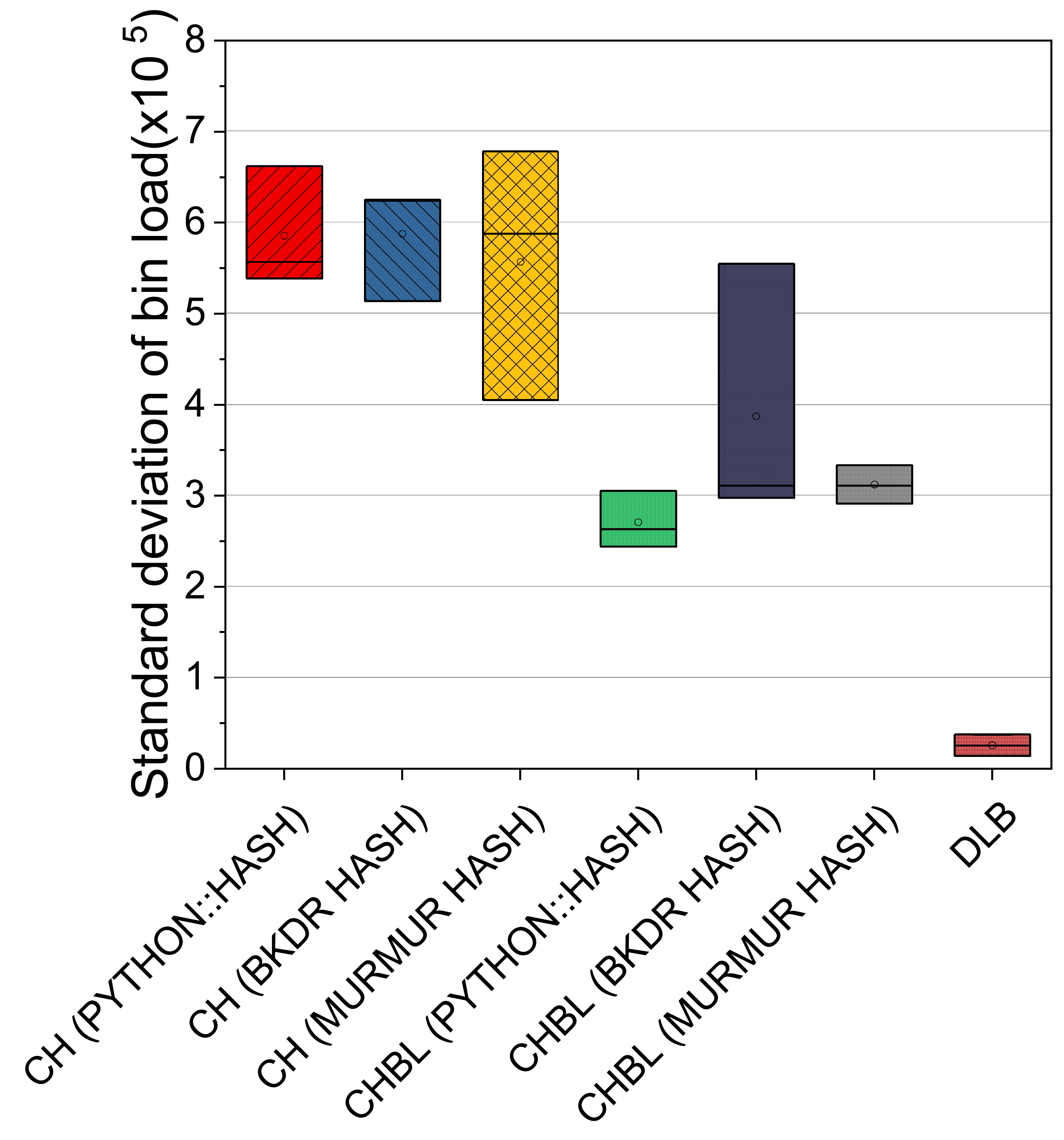}
		\caption{Log-normal}
		\label{fig:analysis:std:log-normal}
	\end{subfigure}
	\begin{subfigure}{.22\textwidth}
		\centering
		\includegraphics[width=1\linewidth]{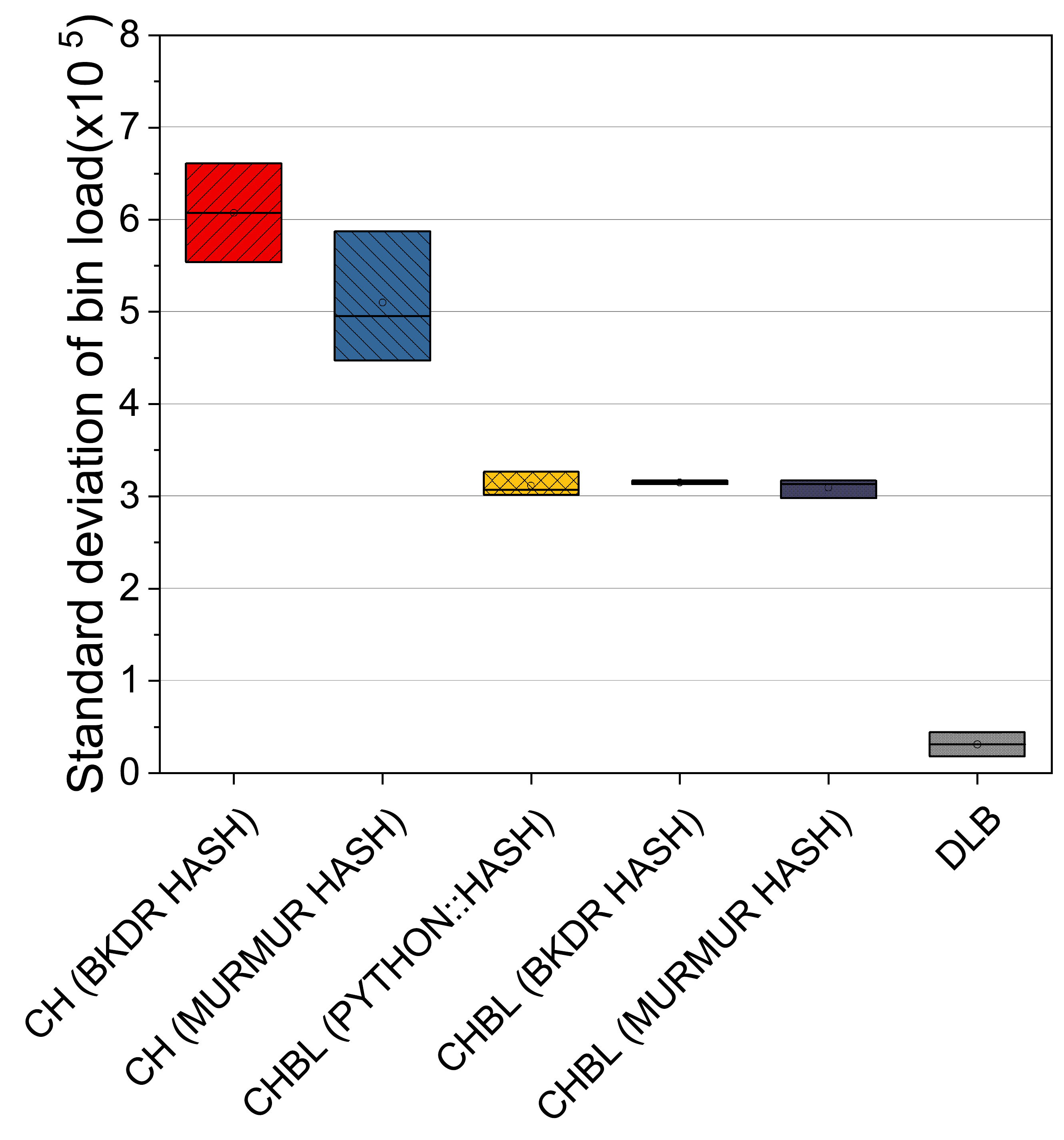}
		\caption{Normal}
		\label{fig:analysis:std:normal}
	\end{subfigure}
	
	\begin{subfigure}{.22\textwidth}
		\centering
		\includegraphics[width=1\linewidth]{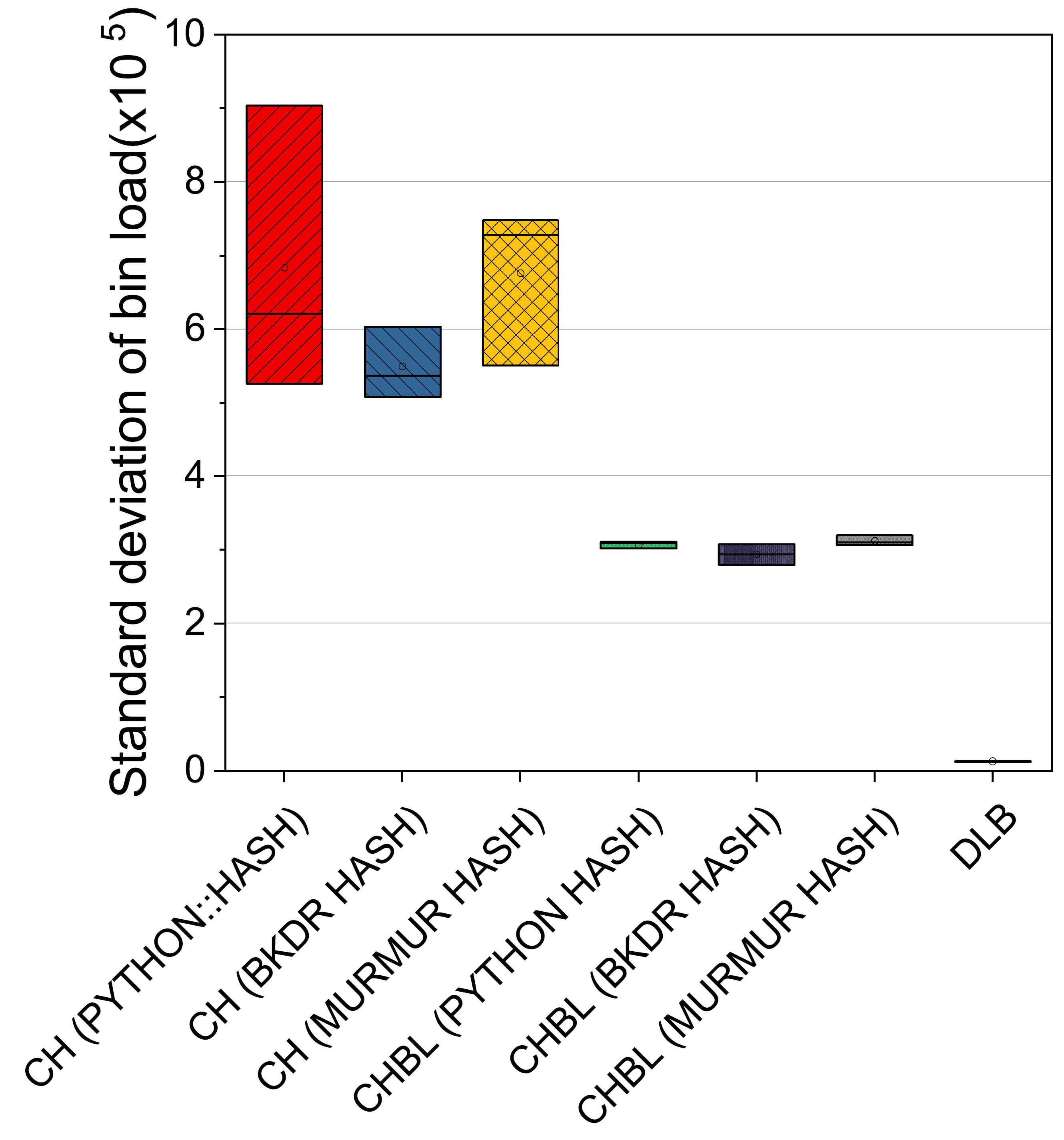}
		\caption{Uniform}
		\label{fig:analysis:std:uniform}
	\end{subfigure}
	\begin{subfigure}{.22\textwidth}
		\centering
		\includegraphics[width=1\linewidth]{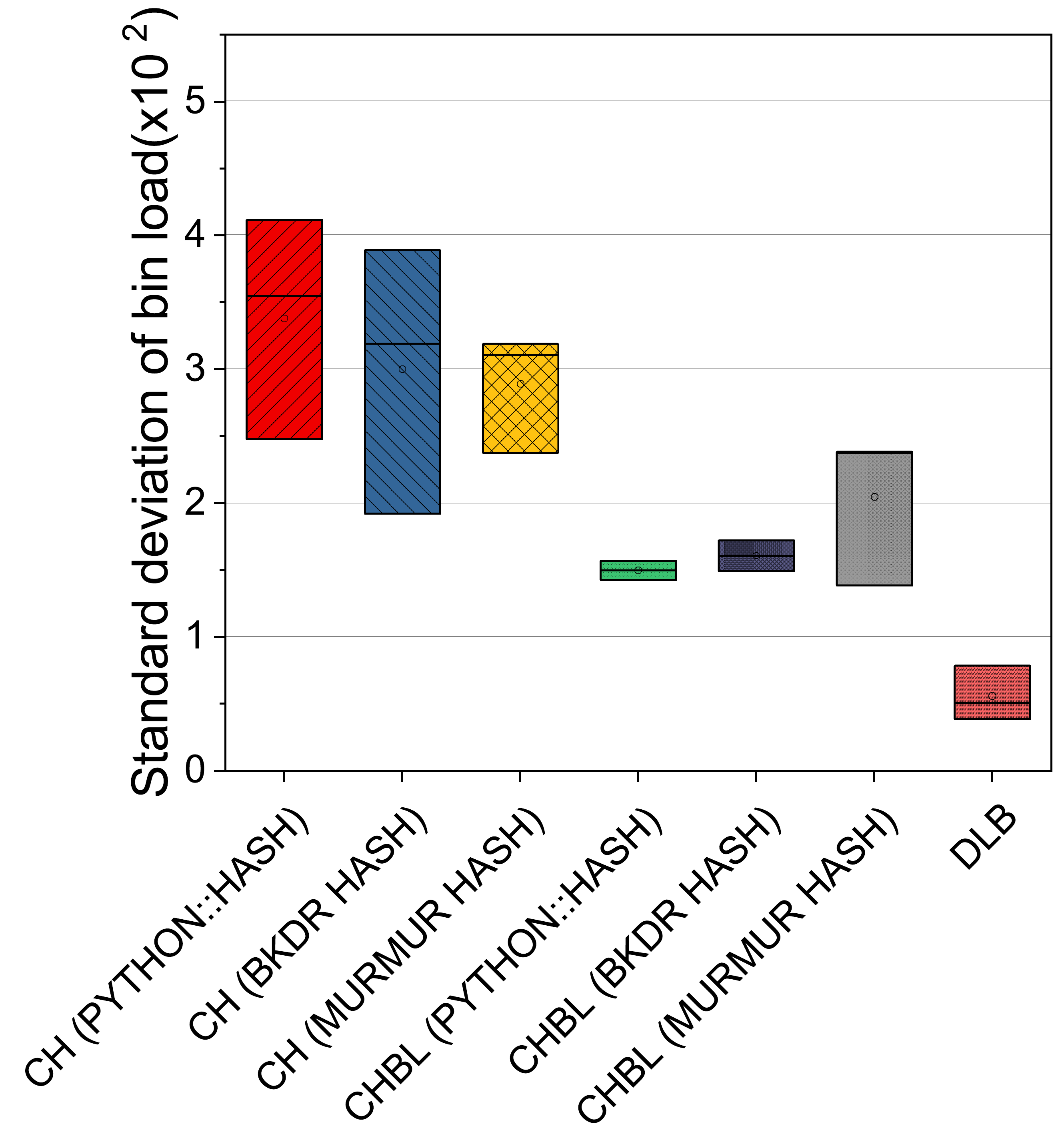}
		\caption{Radio monitoring data}
		\label{fig:analysis:std:real-data}
	\end{subfigure}
	\caption{Standard deviation of bin load using different load balancing mechanisms and distributions of input data sets}
	\label{fig:analysis:std}
\end{figure}
%In this sub-section, we compare the effectiveness of DLB and the other load balancing mechanisms when the input data sets are skewed. 

\subsubsection{Load Balancing}\label{sec:analysis:load-balance}

Figure \ref{fig:analysis:std} compares the $std$ of DLB with CH and CHBL based methods. The $std$ is used to reflect how much an actual distribution of the balls on the hash circle deviates from an idea uniform distribution. For each experiment, we collect the average result as well as the distribution of 10 runs and plot them in Figure \ref{fig:analysis:std}. It can be observed that compared with other methods, DLB has the lowest value of $std$ regardless of the data distributions. For example, when the real-world data set is used, average $std$ value of DLB for the 10 runs is 78, while that of other methods, such as CH(wiht Python Hash) and CH(with BKDR Hash) are 337 and 299 respectively, which are 3.32x and 2.83x larger than that of DLB.

\section{Conclusions}\label{sec:conclusion}
%Load balancing is playing an important role to guarantee the effectiveness of the distributed platforms as well as the performance of various applications and services running on top. 

Existing hash function based load balancing mechanisms cannot perform well when the input data is skewed. In this paper, we proposed DLB, a \underline{D}eep \underline{L}earning based load \underline{B}alancing mechanism, to address this problem. DLB replaces the hash functions in traditional load balancing mechanisms with deep learning models. Given the constant time of model inferencing, using a learned model does not introduce additional runtime overhead compared with using a hash function. %Note that using complex nerual-networks does not mean DLB is ineffective. Actually, since DLB is nerual-network-based(matrix-operations-based), the time complexity of inference is O(1). 
We implemented DLB and deployed it on a practical Cloud environment using CloudSim. Experimental results show that, compared to traditional hash function based load balancing mechanisms, DLB is able to achieve more balanced and stable results even when the input data is skewed. 

%We are also exploring other follow-up directions as future work. One example is to use incremental learning to dynamically adjust the models in load balancing, so that it can work effectively even when the input data distribution varies frequently. Another example is to take advantage of deep learning model to improve the performance of other traditional systems or algorithms.

%Our deepest gratitude goes to the anonymous reviewers for their careful work and thoughtful suggestions that have helped improve this paper substantially. 

%Corresponding author: Wei Zhou (zwei@ynu.edu.cn) and  Jing He(hejing@ynu.edu.cn).

\bibliographystyle{unsrt}  
\bibliography{./references}

\end{document}